\documentclass[twocolumn,showpacs,amssymb,prd]{revtex4}

\usepackage{graphicx}
\usepackage{dcolumn}
\usepackage{bm}
\usepackage{epsfig}

\begin{document}
\newcommand{\gsim}{\hbox{\rlap{$^>$}$_\sim$}}
\newcommand{\lsim}{\hbox{\rlap{$^<$}$_\sim$}}

\title{The Smoking Guns Of Short Hard Gamma Ray Bursts}

\author{Shlomo Dado and Arnon Dar}
\affiliation{Physics Department, Technion, Haifa, Israel}
      
\begin{abstract} 
The X-ray afterglow of short hard bursts (SHBs) of gamma rays 
provides compelling evidence that SHBs are produced by highly 
relativistic jets launched in the birth of rotationally powered 
millisecond pulsars (MSPs) in merger of neutron stars 
and/or  in mass accretion on neutron stars in low 
mass X-ray binaries. Gravitational wave detection of 
relatively nearby neutron star mergers by Ligo-Virgo, followed by
far off-axis short GRBs or orphan afterglows of  
beamed away SHBs with an MSP-like light curve will verify
beyond doubt the neutron star merger origin of SHBs.
\end{abstract}


\maketitle

\section{Introduction} 
Gamma ray bursts seem to be divided into two distinct 
classes, long duration soft gamma ray bursts (GRBs) that usually last 
more than 2 seconds and short hard bursts (SHBs) that usually last less 
than 2 seconds [1]. While there is clear observational evidence 
for production of long duration GRBs  in broad line supernova explosions 
of type Ic [2], the origin of SHBs (and of SN-less GRBs) is still unknown. 
Supernova origin is 
ruled out [3], and although there is no solid evidence, it is widely 
believed [4] that SHBs are produced mainly in merger of neutron stars in 
close binaries [5]. SHBs are often followed by extended emission (EE) 
with a much lower luminosity that lasts a couple of minutes and is taken 
over during its fast decay phase by a long duration afterglow [6]. The origin 
of the EE and the afterglow of SHBs are also not known [4], although it 
has been suggested long ago that rotationally powered 
millisecond pulsars (MSPs) or magnetars 
powered by magnetic energy, which are born in neutron star mergers, 
dominate or contribute to the afterglow of SHBs [7]. Recently, this 
possibility has attracted increasing attention [8], but all past 
attempts to model the X-ray afterglow of SHBs with jet plus 
magnetar emission were based on rather heuristic functions with 
free adjustable parameters rather than on formulae derived
from the underlying model.

In this letter, we show that the X-ray light curves of the 
afterglow of SHBs, which were measured with the Swift X-ray 
telescope (XRT) are indeed those expected from the launch of highly 
relativistic narrowly collimated jets which produce the SHBs in 
the birth of rotationally powered MSPs. 
Such MSPs are the smoking guns from the production of 
SHBs. They seem to be present in all SHBs, which rules out the 
birth of stellar black holes or magnetars powered by the 
decay of  ultra-strong magnetic field  as the origin of 
most SHBs. The highly relativistic jets that produce SHBs in 
neutron star mergers [5,9] or/and in phase transition of 
neutron stars to more compact stars (quark stars ?) in compact 
binaries, within or without globular clusters [10], produce the SHBs 
with or without an EE, respectively.  In SHBs with an EE that are 
produced within dense stellar regions, such as the collapsed cores
of globular clusters, the MSP powered X-ray emission takes over only 
during the fast decay phase of the EE.

\section{Jetted SHBs}
If SHBs, like GRBs, are produced by highly relativistic narrowly 
collimated bipolar jets of plasmoids with a large bulk motion 
Lorentz factor $\gamma>>1$, through inverse Compton 
scattering (ICS) of external light, then they are expected to 
have many similar properties [9,10,11]: Both are narrowly 
beamed, mostly viewed from an angle $\theta\approx 1/\gamma$ relative 
to the jet direction of motion, and
consequently, both are expected to display a large linear polarization, 
$\Pi=2\gamma^2\,\theta^2/(1+\gamma^4\,\theta^4)\approx 1$ [9,11]. 
The peak energy of their time integrated energy spectrum satisfies
$(1+z)\,Ep\propto \gamma\,\delta$, while  
their isotropic equivalent total gamma ray 
energy  satisfies $Eiso\propto \gamma \delta^3$, where $z$ is their 
redshift and $\delta=1/\gamma(1-\beta\, cos\theta)$ is their 
Doppler factor. Hence, ordinary GRBs and SHBs, that are mostly viewed 
from an angle $\theta\approx 1/\gamma$, were predicted to satisfy the 
correlation $(1+z)Ep\propto [Eiso]^{1/2}$, while far off-axis 
($\theta^2>>1/\gamma^2$) GRBs and SHBs were predicted to satisfy 
$(1+z)Ep\propto [Eiso]^{1/3}$ [12]. All these predictions are 
well satisfied by GRBs, but so far, because of observational 
limitations, only the correlation $(1+z)Ep\propto [Eiso]^{1/2}$ for 
near axis (ordinary) SHBs could be verified, as shown in Figure 1. 
\begin{figure}[] 
\centering
\epsfig{file=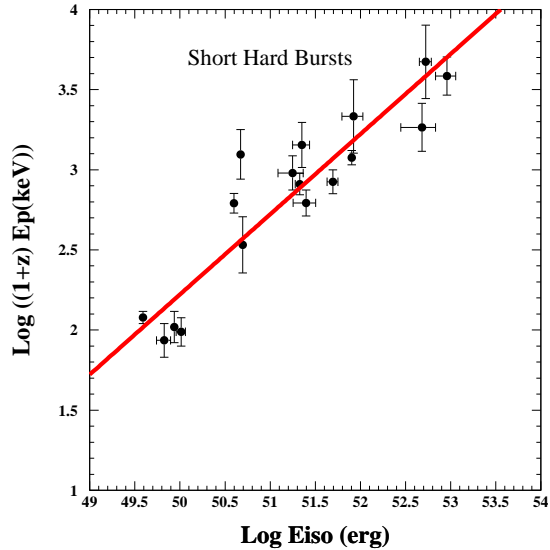,width=8.cm,height=8.cm} 
\caption{The measured rest frame peak energy $(1+z)Ep$ as function of 
the equivalent isotropic  gamma-ray energy $Eiso$ of SHBs with known 
redshift $z$. The line is the expected correlation 
$(1+z)Ep\propto Eiso^{1/2}$ for ordinary (near axis) SHBs 
produced by ICS of light by highly relativistic jets. Far off-axis
short GRBs are expected to satisfy $(1+z)Ep\propto Eiso^{1/3}$.
}  
\label{Fig1}
\end{figure}

\section{Afterglow powered by MSPs}
Young X-ray pulsars seem to be surrounded by nearby plerions,
which absorb their emitted radiation, winds, and highly 
relativistic particles, and convert them to luminous energy, 
mostly in the X-ray band.  Observations indicate that
the change of their period $P$ during their spin down
satisfies to a good approximation,
\begin{equation}
P\dot{P}=K
\label{Eq.1}
\end{equation}
where $K$ is time independent constant. Such a behavior is 
expected from pulsars which spin down in vacuum by magnetic 
dipole radiation (MDR) [13]. 
But  $P\,\dot{P}$ seems to remain constant to a good 
approximation also when other spin down mechanisms, such as 
emission of winds and/or highly relativistic particles, 
contribute, or even dominate the spin-down of pulsars. This 
is suggested by the fact that the age estimate 
$t=[P^2(t)-P^2(0)]/2\,K$ and the braking relation 
$d^2P/dt^2\approx -K^2/P^3$ [13] 
seem to be well satisfied by young X-ray pulsars: The ages 
obtained from the age relation are consistent with
the known ages of their parent 
supernovae (historical supernova)  and/or their ages obtained from 
measurements of the distance and proper motion of the young 
pulsars relative to the centers of the supernova remnants where 
they were born. 
The braking relation, which is very difficult to test
over a human time scale, has been verified only in few young
X-ray pulsars where it yielded a braking index near the expected
value 3 [14].

Eq.(1) yields a time-dependent period 
\begin{equation} 
P(t)=P_i(1 +t/t_b)^{1/2}, 
\label{Eq2} 
\end{equation} 
where $P_i=P(0)$ is the initial period of the neutron star,
$t$ is the time after its birth, and  $t_b=P_i/2\,\dot{P}_i$. 

If the rotational energy loss of a neutron star with a constant 
moment of inertia $I$  powers its luminosity $L$ and 
if $P\dot{P}$ remains constant as function of time, then its  
luminosity 
$L=I\omega\, \dot{\omega}$ where $\omega=2\, \pi/P$ satisfies
\begin{equation}
L_{ps}(t)=L_{ps}(0)\,(1 + t/t_b)^{-2}.
\label{Eq3}  
\end{equation}

The afterglow of a
GRB at redshift $z$  that is powered by such a luminosity has a 
local energy flux density $F(t)=L(t)/4\,\pi\ D_L^2$, where $D_L$ 
is the luminosity distance of the GRB at redshift $z$, i.e.,
\begin{equation}
F_{ps}(t)=F_{ps}(0)\,(1 + t/t_b)^{-2}\, .
\label{Eq4}
\end{equation}
If only a fraction $\eta$ of the rotational energy loss is 
converted to X-rays, then
\begin{equation}
P_i={1\over D_L}\sqrt{{(1+z)\,\pi\, \eta_x\, I \over  
2\,F_{ps,x}(0)\,t_b}},
\label{Eq5}  
\end{equation}
For a  canonical neutron star of a  radius 
$R=10^6$ cm and a mass $M\approx M_{Ch}$, where
$M_{Ch}\approx 1.4 M_\odot$ is the Chandrasekhar mass limit of 
white dwarfs, the  moment of inertia has the value
$I=(2/5)M_{Ch} R^2\approx 1.12\times 10^{45} g\, cm^2$.

\section{Light curves powered by magnetars}

The standard estimate [13] of the magnetic field of pulsars is valid 
only for MSPs in vacuum, which spin down by emitting magnetic dipole 
radiation (MDR). It 
overestimate the magnetic field if other power sources dominate 
their electromagnetic radiation and is unreliable for distinguishing 
between highly magnetized MSPs and magnetars. Therefore we shall 
adhere to the original definition of magnetars, namely pulsars whose 
luminosity is powered by their ultra strong magnetic field energy, 
rather than by their rotational energy or other intrinsic power 
sources.

A magnetic dipole ${\bf m}$ of a neutron star of a radius $R$
produces a magnetic field whose peak surface value $B=2\,m/R^3$ 
is at its magnetic poles, and its total magnetic field energy is
$U\approx B^2 R^3/12$. If the magnetic dipole is aligned at an 
angle $\alpha$ relative to the rotation axis of the neutron star 
whose spin period is $P$, then it emits MDR at a rate [15],
\begin{equation}
L={8\,\pi^4\,B^2\,R^6\, sin^2\alpha\over 3\,c^3\,P^4}\,.
\label{Eq6}
\end{equation}
However, since both $L$ 
and $U$ are proportional to $B^2$, the
observed luminosity of such  magnetars decreases like 
\begin{equation}
L(t)= L(0)(P_i/P)^4\,exp(-\int dt/\tau)
\label{Eq7}
\end{equation}
where  $\tau=c^3\,P^4/64\,\pi^4\,R^3\,sin^2\alpha$.
As long as $P(t)\approx P_i$,
\begin{equation}
L(t)= L(0)\, exp(-t/\tau))\, . 
\label{Eq8}
\end{equation}
The different light-curves of MSPs powered by rotation energy  
and of magnetars powered by magnetic field energy, as given, respectively, 
by Eq.(4) and Eq.(8), can tell 
which one of these sources, if any, can reproduce the light-curves 
of the X-ray afterglows of SHBs  and  be identified as the 
smoking gun of SHBs.

Figures 2,3 show the well sampled X-ray afterglow of two 
representative SHBs without EE, 130603B and 090510, which 
were reported in the Swift-XRT GRB light curve 
repository [16] and their best fit MSP and magnetar 
powered light curves as given, respectively, by Eq.(4) and 
Eq.(8). 
\begin{figure}[]
\centering
\epsfig{file=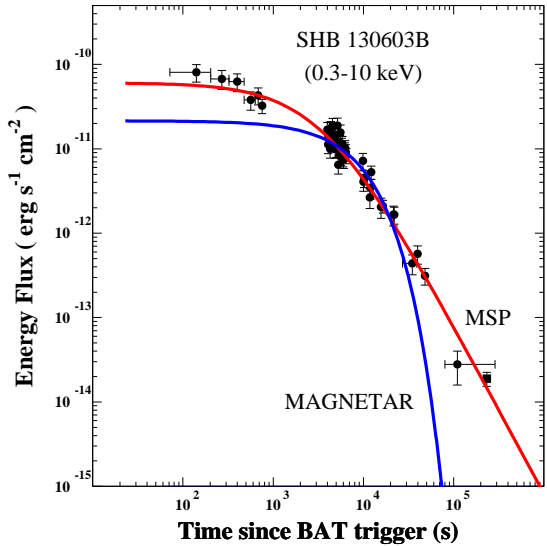,width=8.cm,height=8.cm}
\caption{The light curve of the X-ray afterglow of SHB 130603B 
reported in the Swift-XRT GRB light curve repository [16] and  
its best fit light curves of MSP powered by rotation
or by a magnetar powered by magnetic field energy, 
as given by Eqs. (4) and (8), respectively, with the parameters listed in
table I.}
\label{Fig2}   
\end{figure}
\begin{figure}[] 
\centering
\epsfig{file=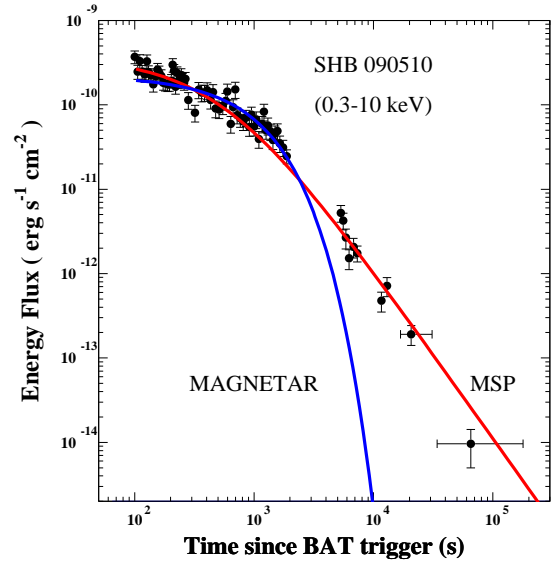,width=8.cm,height=8.cm}
\caption{The light curve of the X-ray afterglow of SHB 090510
reported in the Swift-XRT GRB light curve repository [16] and
its best fit light curves of MSP powered by rotation 
or by a magnetar powered by magnetic field energy,              
as given by Eqs. (4) and (8), respectively, with the parameters listed in 
table I.}
\label{Fig3}
\end{figure} 
As demonstrated in figures 2,3, while Eq.(4) for rotationally 
powered MSPs fit well the X-ray afterglows of SHBs 130603B and 
090510 without EE [16], the magnetar light curves, which are 
powered by magnetic field energy, as given by Eq.(8), do not. 
This is valid  for all the well sampled X-ray afterglows 
of SHBs without EE that are reported in the Swift-XRT GRB 
light curve repository [16].

\section{SHBs with an extended emission.} 
A considerable fraction of SHBs may take place in rich star clusters, such 
as globular clusters (GCs) [10], in particular within collapsed cores (CCs) 
of GCs where the ratio of binary neutron stars and MSPs to ordinary stars is 
much higher than in the regular interstellar medium of galaxies. ICS of the 
ambient light around the birth place of the MSP within the GCs by the highly 
relativistic jet/plasmoid, which produces the SHB, can also produce the 
extended emission (EE). Inside the GC where the light is nearly isotropic, 
the EE is proportional to the light density along the plasmoid's trajectory. 
Far outside the GC, the intercepted light has a small angle relative to the 
plasmoid direction of motion. At a distance $r$ from a GC of a radius $R$, 
the energy $\epsilon_\gamma$ of the intercepted photons is Lorentz transformed 
to $\approx (R/r)\epsilon_\gamma$ in the jet/plasmoid rest frame. This 
energy is boosted by ICS to an energy $E\approx \gamma\,\epsilon_\gamma \, 
R/(1+z)\,r$ when observed at a viewing angle $\theta\approx 1/\gamma$ in the 
rest frame of the distant observer. At observer time $t$, 
$r=\gamma\delta\,c\,t/(1+z)$. Hence, for a GC light with bremsstrahlung-like 
spectrum $\epsilon^2\, dn/d\epsilon \propto exp(-\epsilon/\epsilon_c)$ and a 
photon density $n_\gamma \propto 1/r^2$, the observed energy flux density of 
the EE due to ICS of GC light decays like 
\begin{equation} 
F_{ee}(t)\propto  t^{-1} exp(-t/t_{ee})
\label{Eq9}
\end{equation}
with $t_{ee}(E)\approx 100\,(R/pc)(\epsilon_c/eV)/c\,(E/keV)(\gamma/10^3)$s, 
which depends strongly on the observed energy band, and produces a
fast spectral softening of the EE during its rapid decay phase.
A succession of plasmoids or a non homogenous light density within 
the star cluster yield a bumpy EE light curve,
which can be resolved because of the relatively large photon 
statistics.  Later, when the luminosity decreases, 
the larger time bins (and perhaps merger of  plasmoids)    
yield a smoothly appearing light curves. Hence, the smoothed X-ray 
light-curves of SHBs with EE, which are taken over by MSP,  
are expected to be described roughly by 
\begin{equation}
 F(t)\approx {F_{ee}\,a_{ee}\over a_{ee}+ (t/t_{ee})exp(t/t_{ee})}+
{F_p\over (1+t/t_b)^2}\,,
\label{Eq10}
\end{equation}
while those taken over by magnetars are expected to be described 
roughly by Eq.(10) with $1/(1+t/t_b)^2$  replaced by 
$exp(-t/\tau)$. In Eq. (10), the  parameter $a_{ee}$  was 
introduced for the interpolation between a smoothed  early time 
EE behavior and its late time analytic description (Eq. (9)). 

\section{Comparison with observations}
Figures 4,5 show the X-ray afterglow of two representative 
SHBs with EE, 150424A and 060313, with well sampled light curves
reported in the Swift-XRT GRB light curve repository [16].
Also shown are their best fit light curves obtained
from Eq.(10) with essentially 4 adjustable parameters 
(2 for the decline of EE and 2 for the MSP or magnetar contribution).
\begin{figure}[]
\centering
\epsfig{file=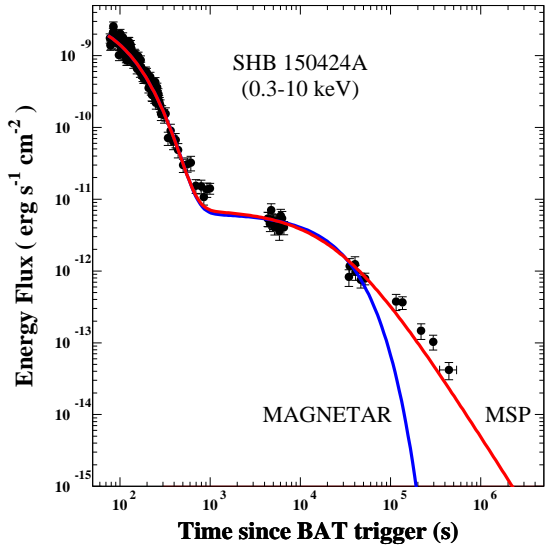,width=8.cm,height=8.cm}    
\caption{The light curve of the X-ray afterglow of SHB 150424A
reported in the Swift-XRT GRB light curve repository [16] and
its best fit light curves of MSP powered by rotation
or by a magnetar powered by magnetic field energy,
as given by Eqs. (4) and (8), respectively, with the parameters listed in
table I.}
\label{Fig4}
\end{figure} 
\begin{figure}[]
\centering
\epsfig{file=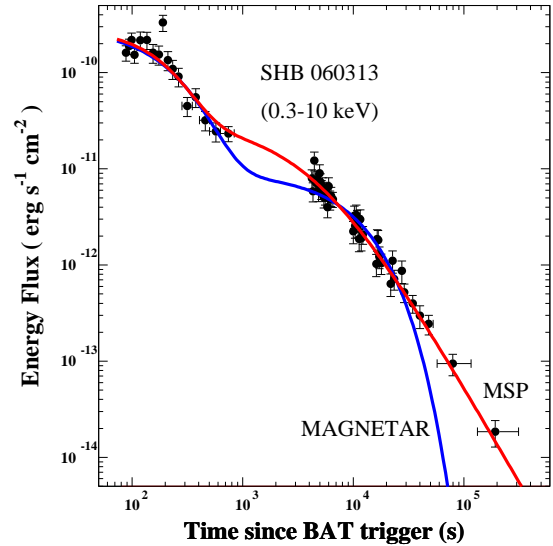,width=8.cm,height=8.cm}    
\caption{The light curve of the X-ray afterglow of SHB 060313
reported in the Swift-XRT GRB light curve repository [16] and
its best fit light curves of MSP powered by rotation
or by a magnetar powered by magnetic field energy,
as given by Eqs. (4) and (8), respectively, with the parameters listed in
table I.}
\label{Fig5}
\end{figure}
As shown in figures 2-5 and in figures 6-9 in the appendix, 
Eq.(10) fits well the X-ray lightcurves of 
all the 20 well sampled X-ray afterglows 
of bursts  with a secured SHB identity,
which are reported todate in the Swift-XRT GRB light curve 
repository [16]. The best fit parameters are 
listed in Table I. 

\noindent
{\bf Conclusions:}  
All the well sampled X-ray afterglow of SHBs [16] 
provide compelling evidence that most SHBs are produced by highly 
relativistic jets which are launched in the birth of rotationally 
powered MSPs, and not in the birth of stellar black holes or 
magnetars powered by magnetic field energy: 

The observed $Ep\,-\,Eiso$ correlation shown in Figure 1 
for near axis SHBs is that expected from ICS of light by highly 
relativistic jets. The correlation $(1+z)\, Ep\propto Eiso^{1/3}$ 
for far off-axis SHBs could not be verified because of a small 
sample of such nearby SHBs.
The observed fast decline phase of the EE with a rapid spectral 
softening [16] is well predicted by Eq.(9). The observed  
durations
$t_{ee}$ of the extended emission of SHBs are compatible with the 
typical size of the collapsed cores of rich globular clusters, which 
are the most likely sites 
of neutron star mergers. Measurements of the polarization of 
the prompt $\gamma$-rays and the EE of bright SHBs, with 
space based polarimeters such as Astrosat [17] and/or Polar 
[18] may provide a critical test of jet production 
mechanism of SHBs, which predicts a linear polarization, 
$\Pi\sim 1$ for near axis SHBs and their EE that declines 
to $\Pi\sim 0 $ during the EE rapid decline phase.

MSP powered light curves fit very well the observed late-time 
X-ray afterglow of SHBs. The MSP periods $Pi$ obtained from 
Eq.(6) are all well above the classical lower bound 
$2\,\pi\, R/c > 0.2\, ms$ for canonical neutron stars with a 
radius $R\approx 10^6\, cm$ (or of quark stars with smaller 
radii).  The small periods are  expected from 
the birth of MSPs in neutron star mergers or in 
the collapse of neutron stars to quark stars due to mass accretion 
in compact binaries [10].

The jet contribution to the observed prompt emission and EE phases 
overshines the MSP contribution during these phases. Absorption and 
re-emission of the MSP spin down energy  by plerions probably turn 
a direct detection of a periodic MSP signal during the prompt and 
extended emission phases [19] to almost an impossible mission. 
However, the quasi-isotropic emission 
powered by the MSP is expected to be the only visible emission from 
far-off axis SHBs. Such emission from nearby MSPs may be detected 
for days, months, and perhaps even years after their birth, and may 
explain the origin of some ultraluminous X-ray (ULX) sources [20]. 
The much smaller detection horizon of such orphan afterglows compared to 
that of near-axis SHBs, is over compensated by not being beamed. 
Gravitational wave detection of relatively nearby ($D_L< 100$ Mpc)
neutron star mergers by Ligo-Virgo, followed by the detection of 
far off-axis SHBs  or an orphan afterglow of a
beamed away SHB with an MSP-like light curve will verify 
the neutron star merger origin of SHBs.

\appendix

\begin{figure}[]
\centering
\vspace{-1cm}
\vbox{
\hbox{
\epsfig{file=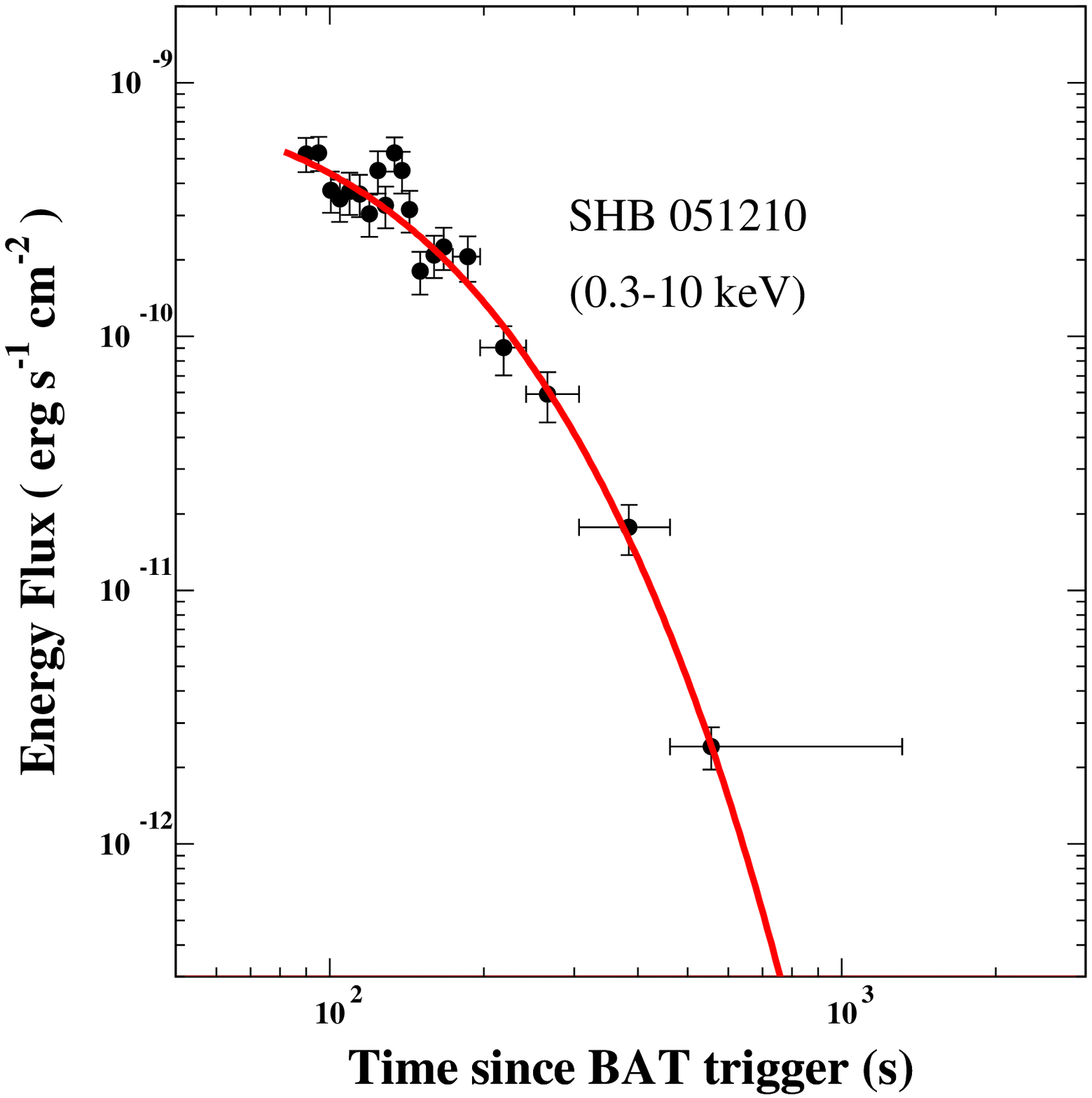,width=8.6cm,height=8.0cm}
\epsfig{file=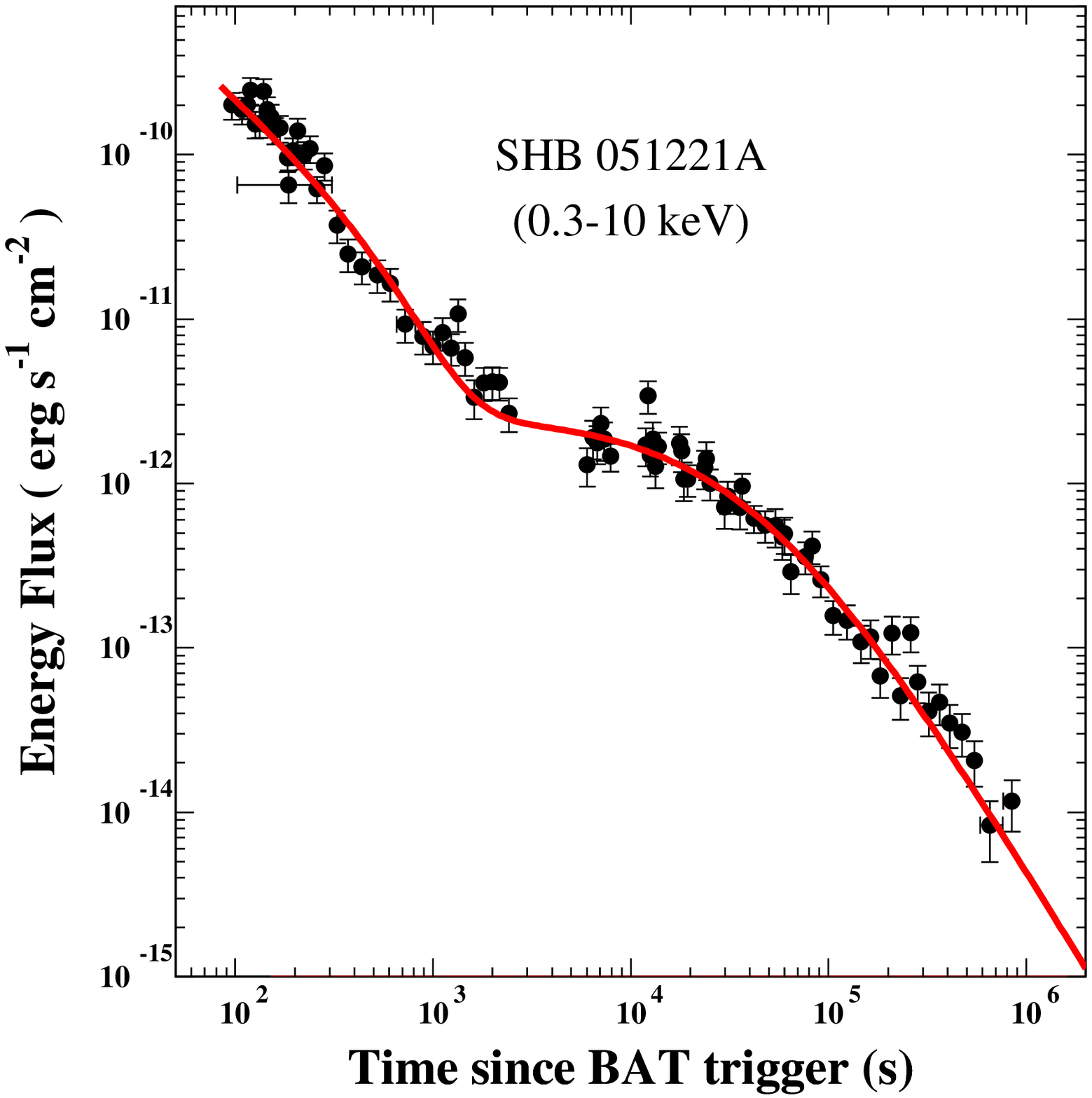,width=8.6cm,height=8.0cm}
}}
\vbox{
\hbox{
\epsfig{file=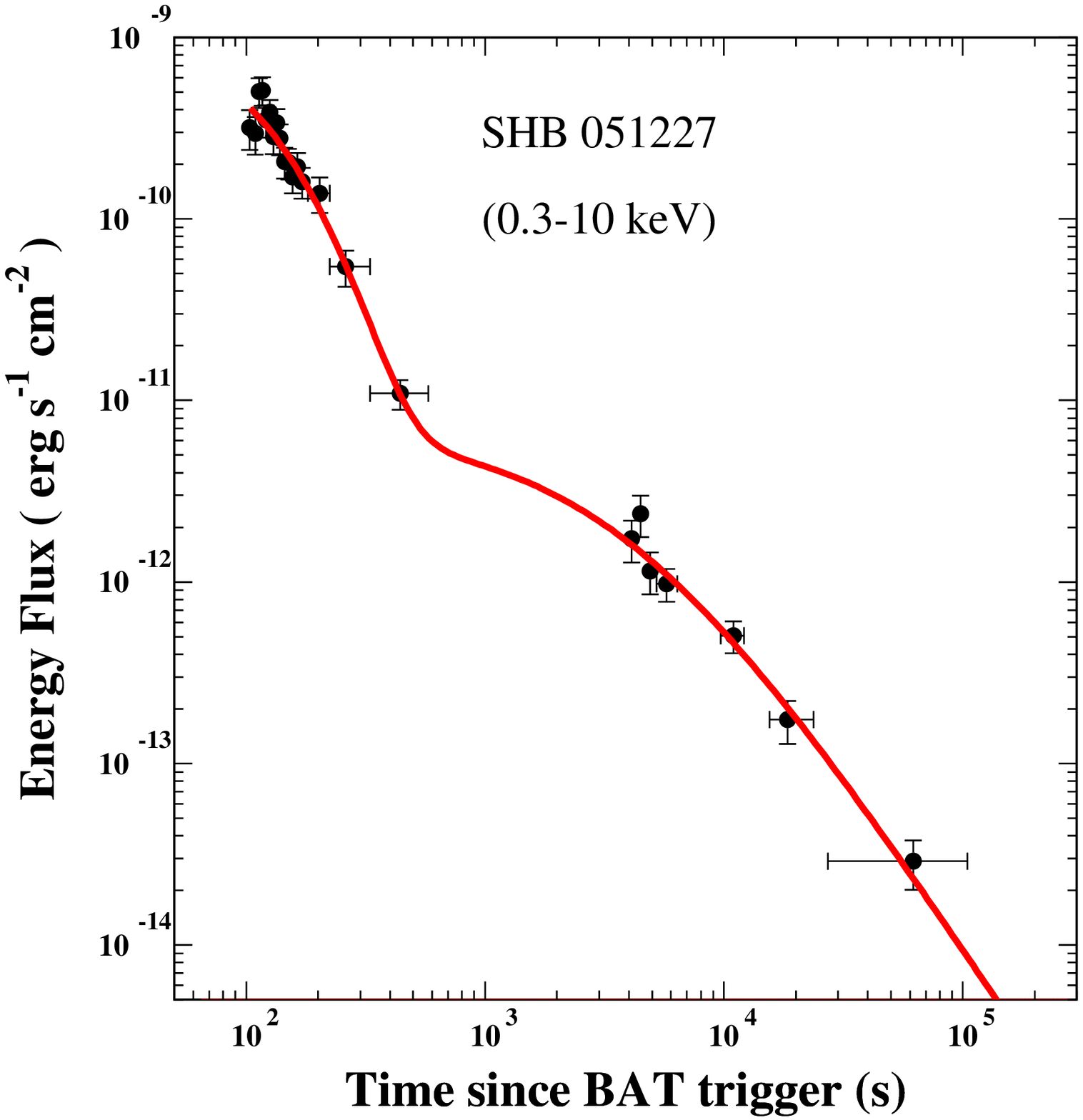,width=8.6cm,height=8.0cm}
\epsfig{file=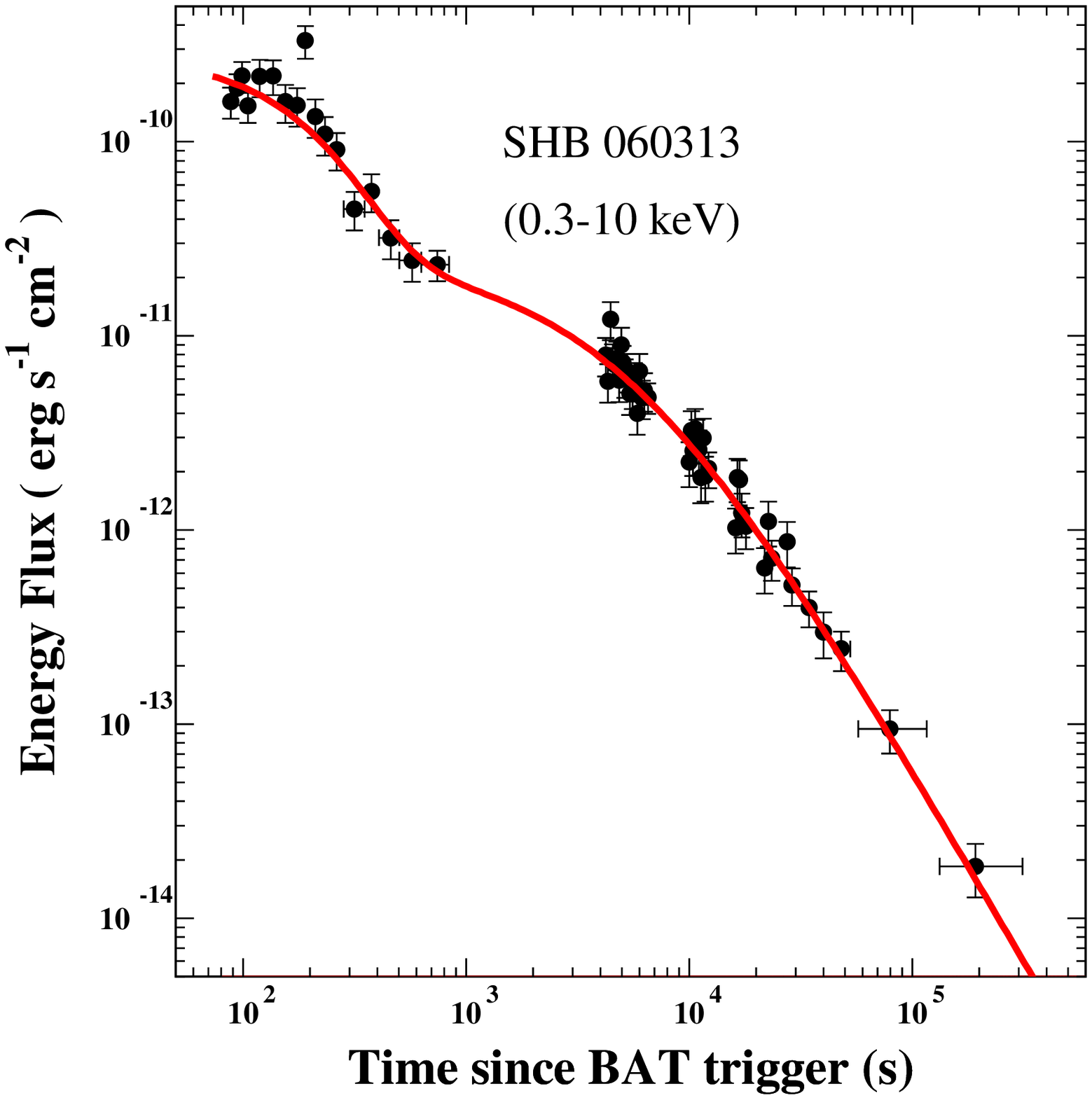,width=8.6cm,height=8.0cm}
}}
\vbox{
\hbox{
\epsfig{file=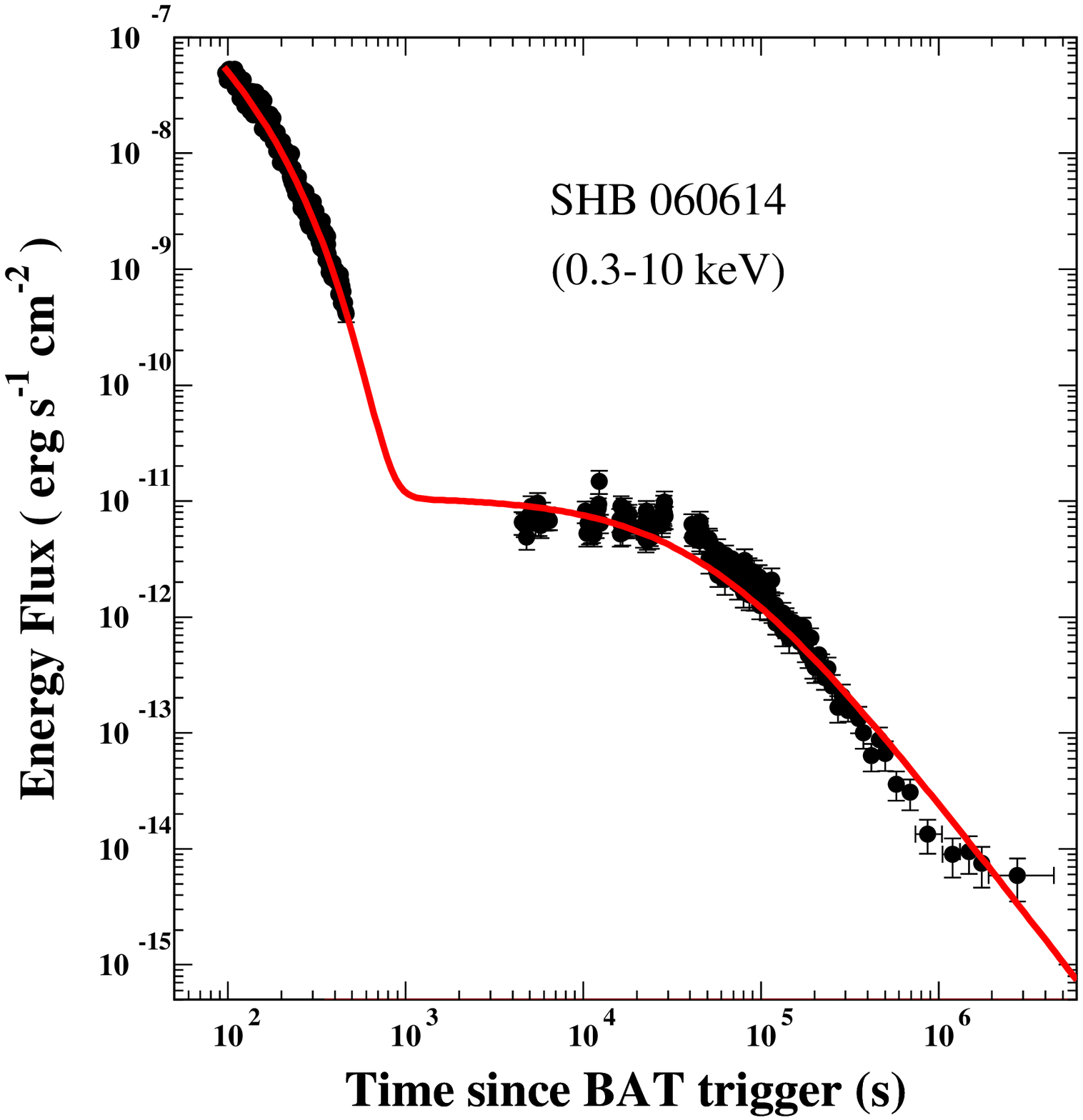,width=8.6cm,height=8.0cm}
\epsfig{file=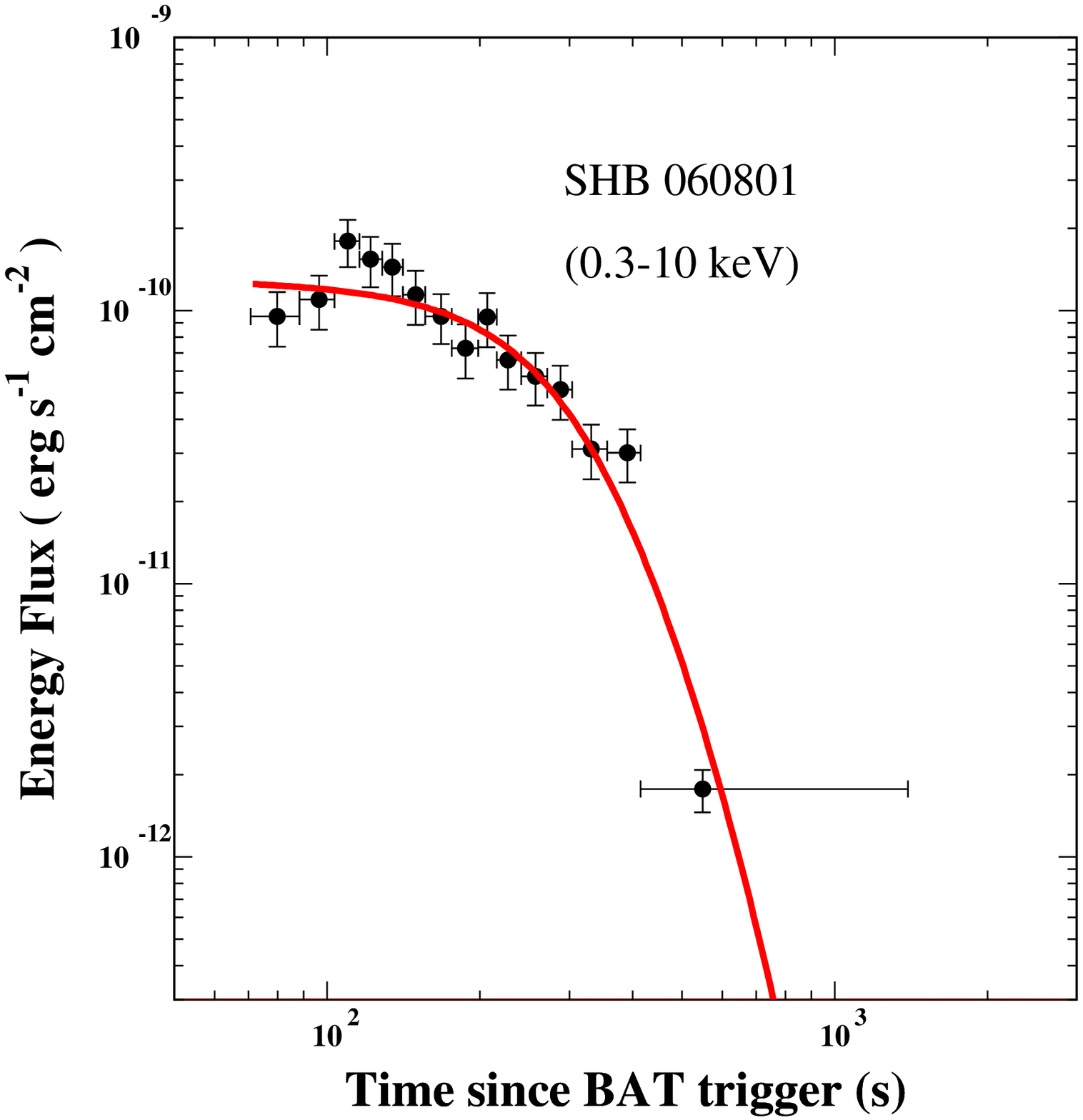,width=8.6cm,height=8.0cm}
}}
\caption{The X-ray light curve of  SHBs
051210, 051221A, 051227, 060313, 060614, and 060801
reported in the Swift-XRT GRB light curve repository [16]
and their best fit light curves assuming ICS of GC light by a 
highly relativistic jet  taken  over
by MSP afterglow, as given by by Eq.(4) or Eq.(10)  
with the 2 or 4 parameters, respectively, listed in Table I.}
\label{F6}
\end{figure}

\begin{figure}[]
\centering
\vspace{-1cm}
\vbox{
\hbox{
\epsfig{file=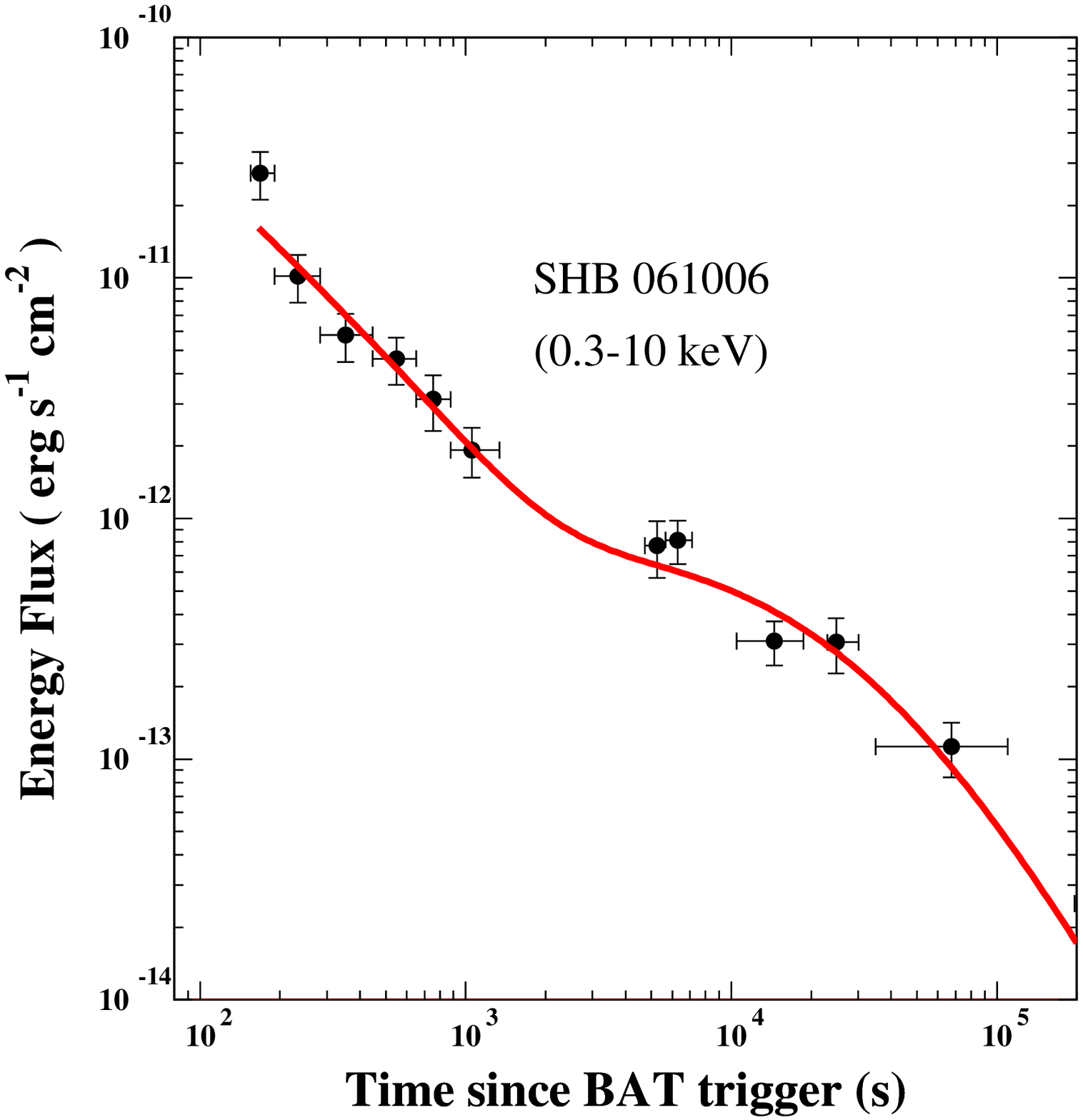,width=8.6cm,height=8.0cm}
\epsfig{file=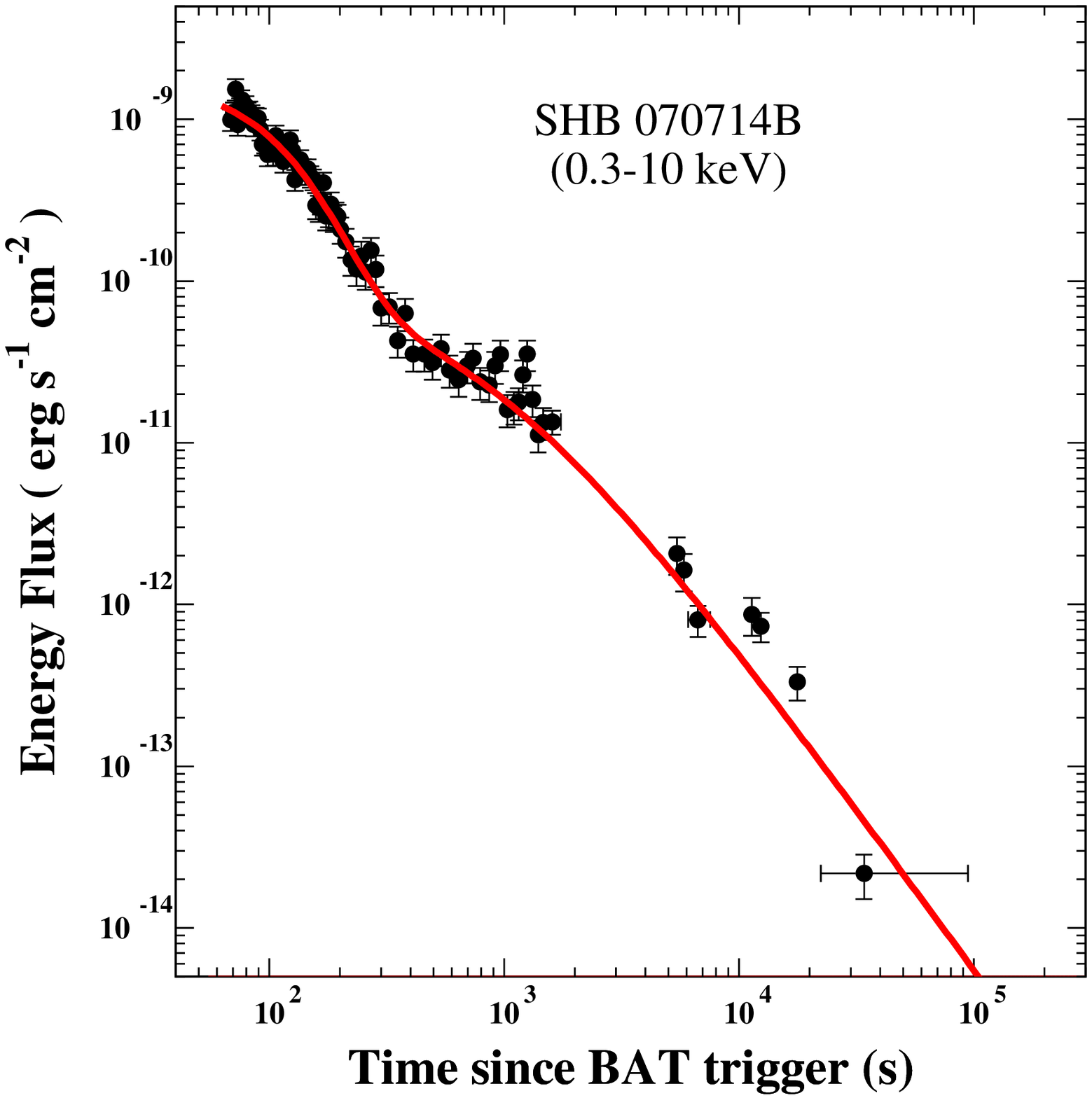,width=8.6cm,height=8.0cm} 
}}
\vbox{
\hbox{      
\epsfig{file=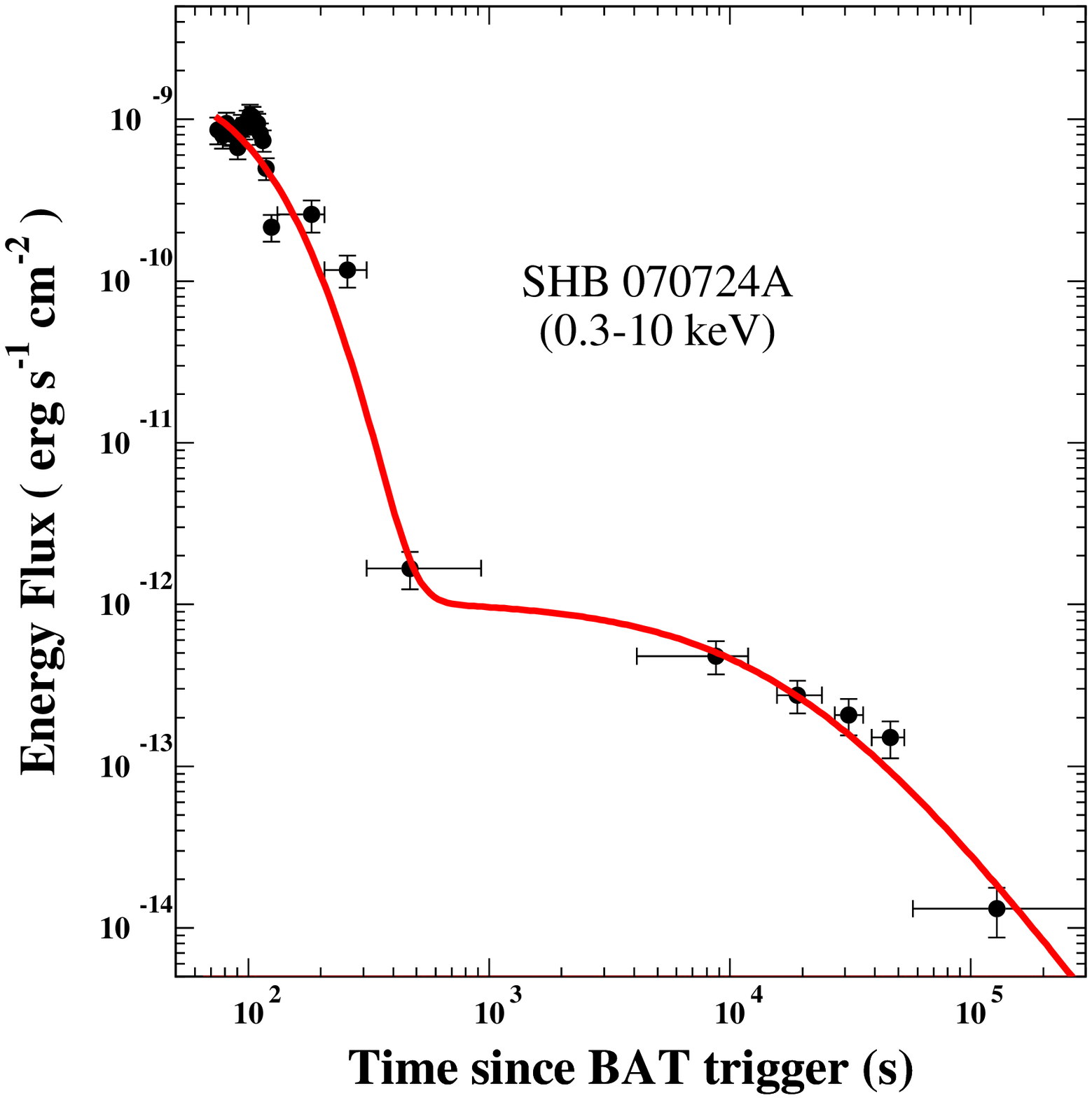,width=8.6cm,height=8.0cm}
\epsfig{file=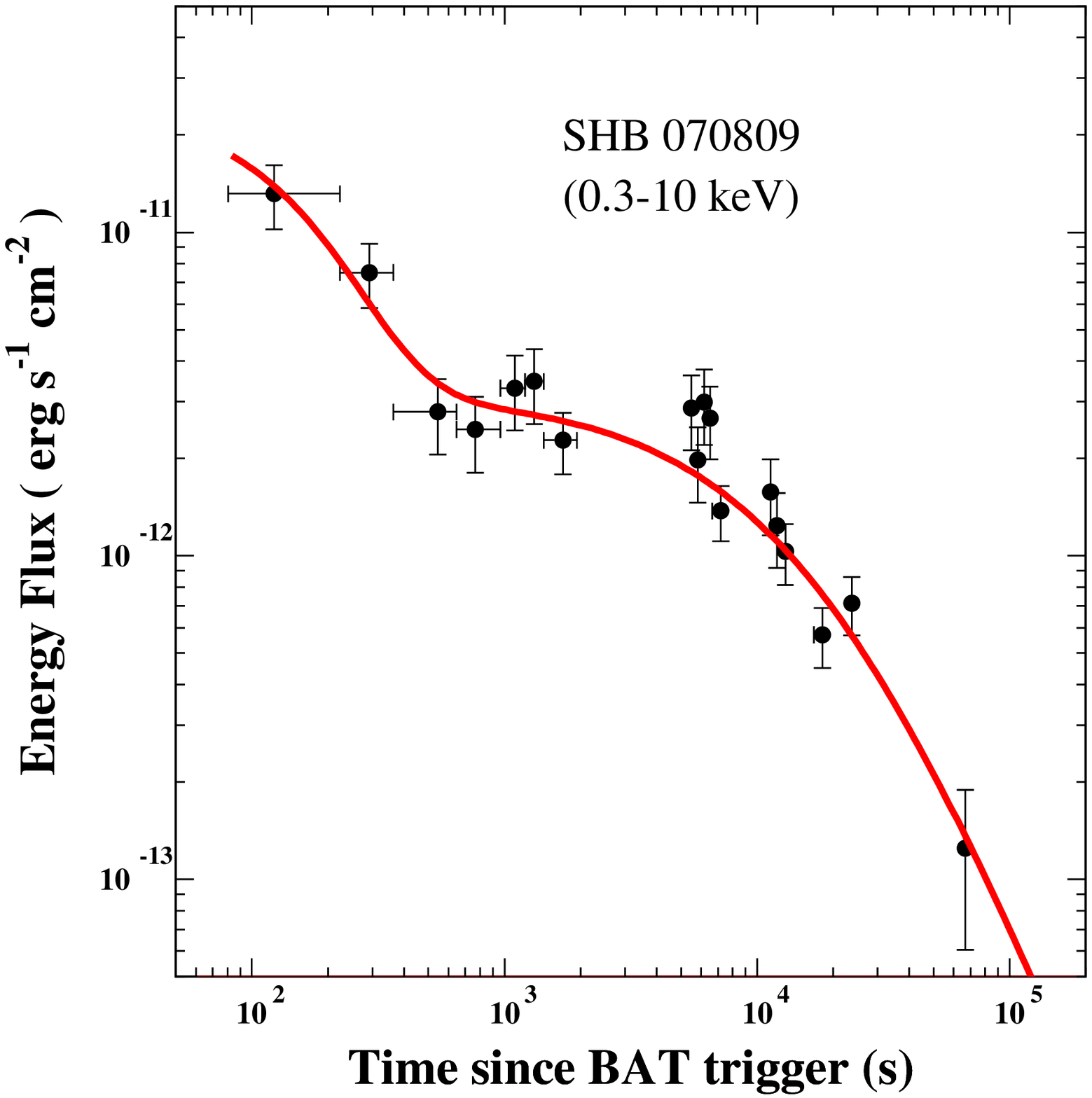,width=8.6cm,height=8.0cm}
}}
\vbox{
\hbox{
\epsfig{file=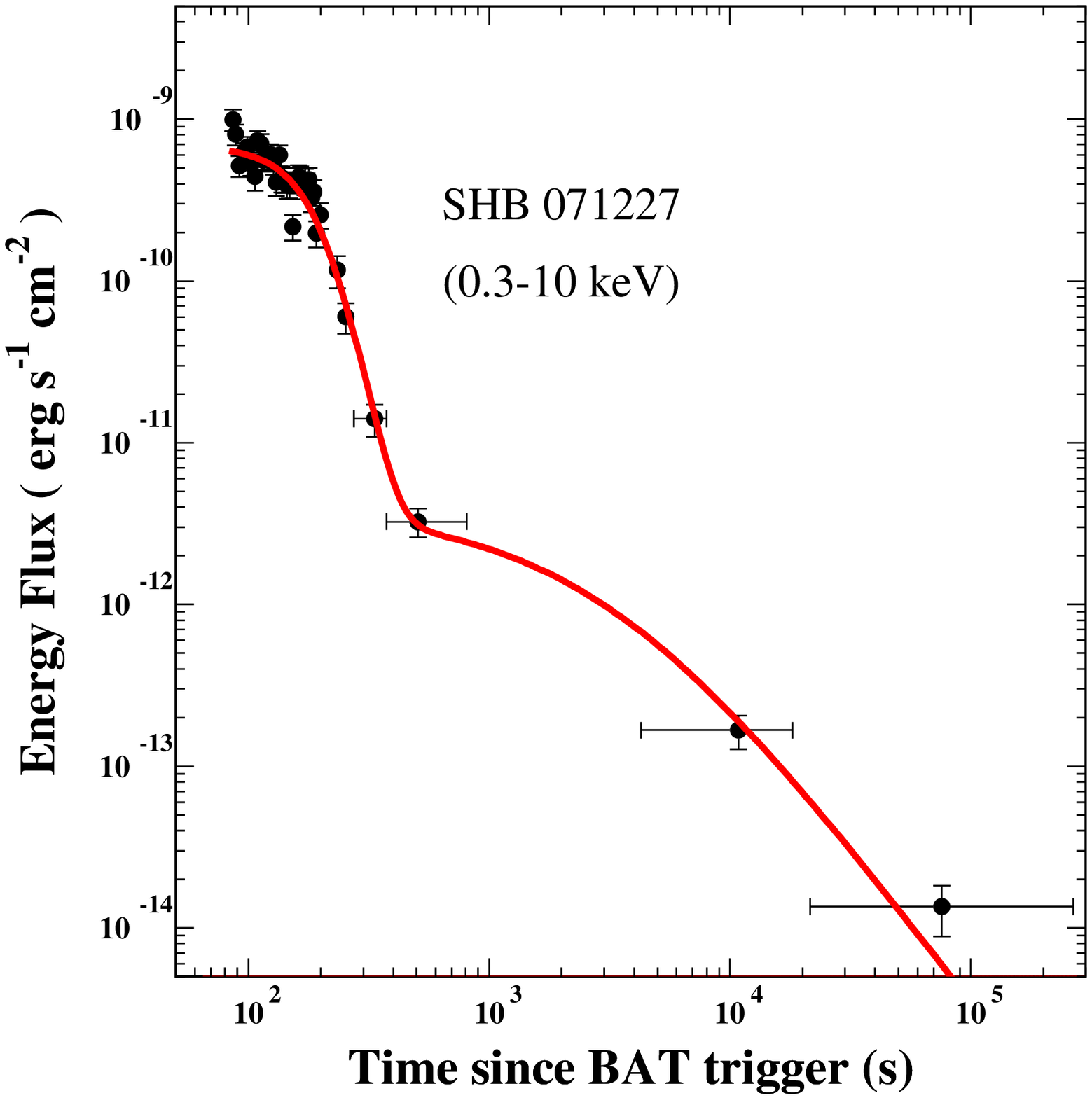,width=8.6cm,height=8.0cm}
\epsfig{file=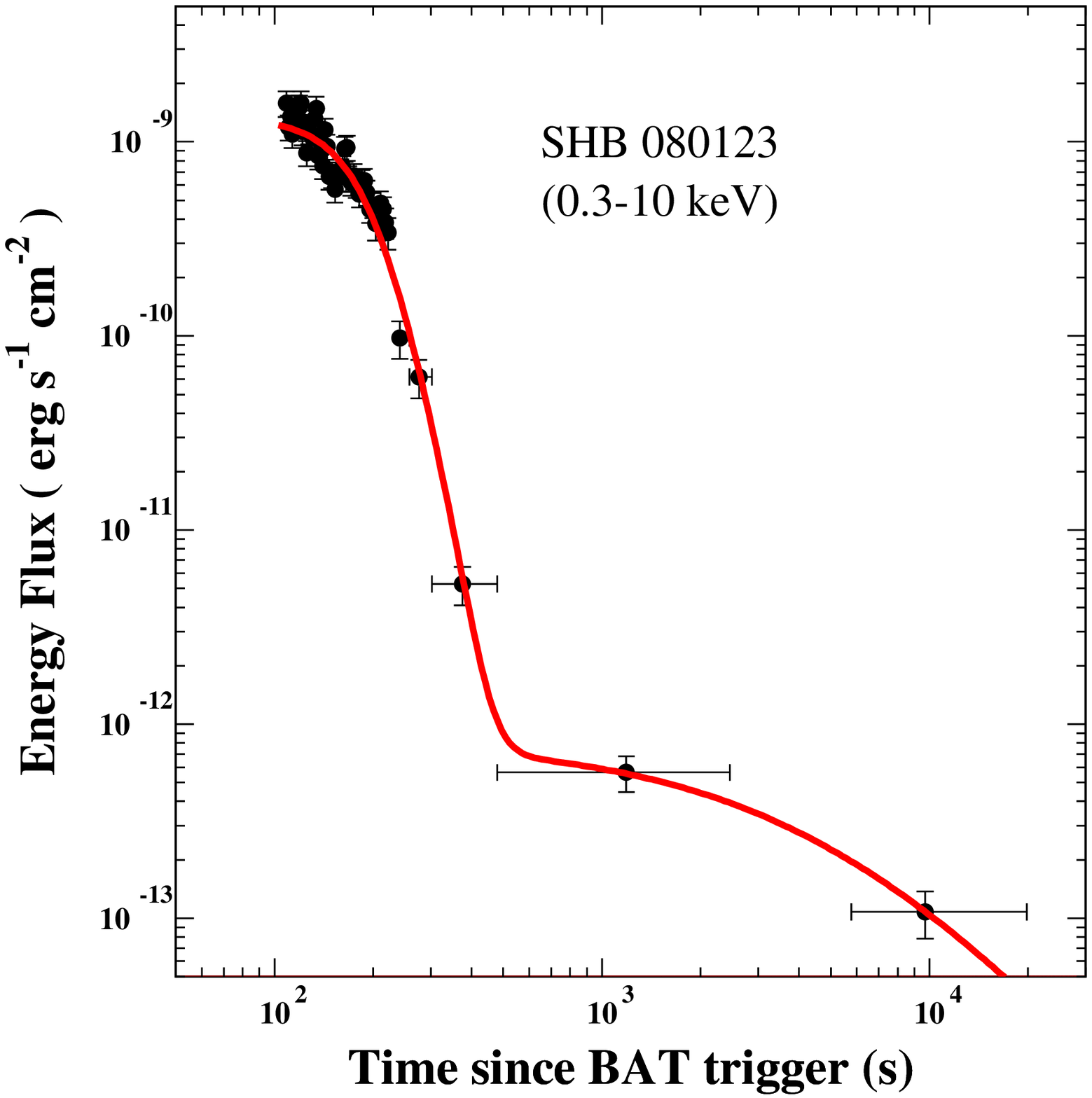,width=8.6cm,height=8.0cm}
}}
\caption{The X-ray light curve of  SHBs
061006, 070714B, 070724A, 070809, 071227, and  080123
reported in the Swift-XRT GRB light curve repository [16]
and their best fit light curves assuming  ICS of GC light by a
highly relativistic jet  taken  over
by MSP afterglow, as given by by Eq.(4) or Eq.(10)
with the 2 or 4 parameters, respectively, listed in Table I.}
\label{F7}
\end{figure}

\begin{figure}[]
\centering
\vspace{-1cm}
\vbox{
\hbox{
\epsfig{file=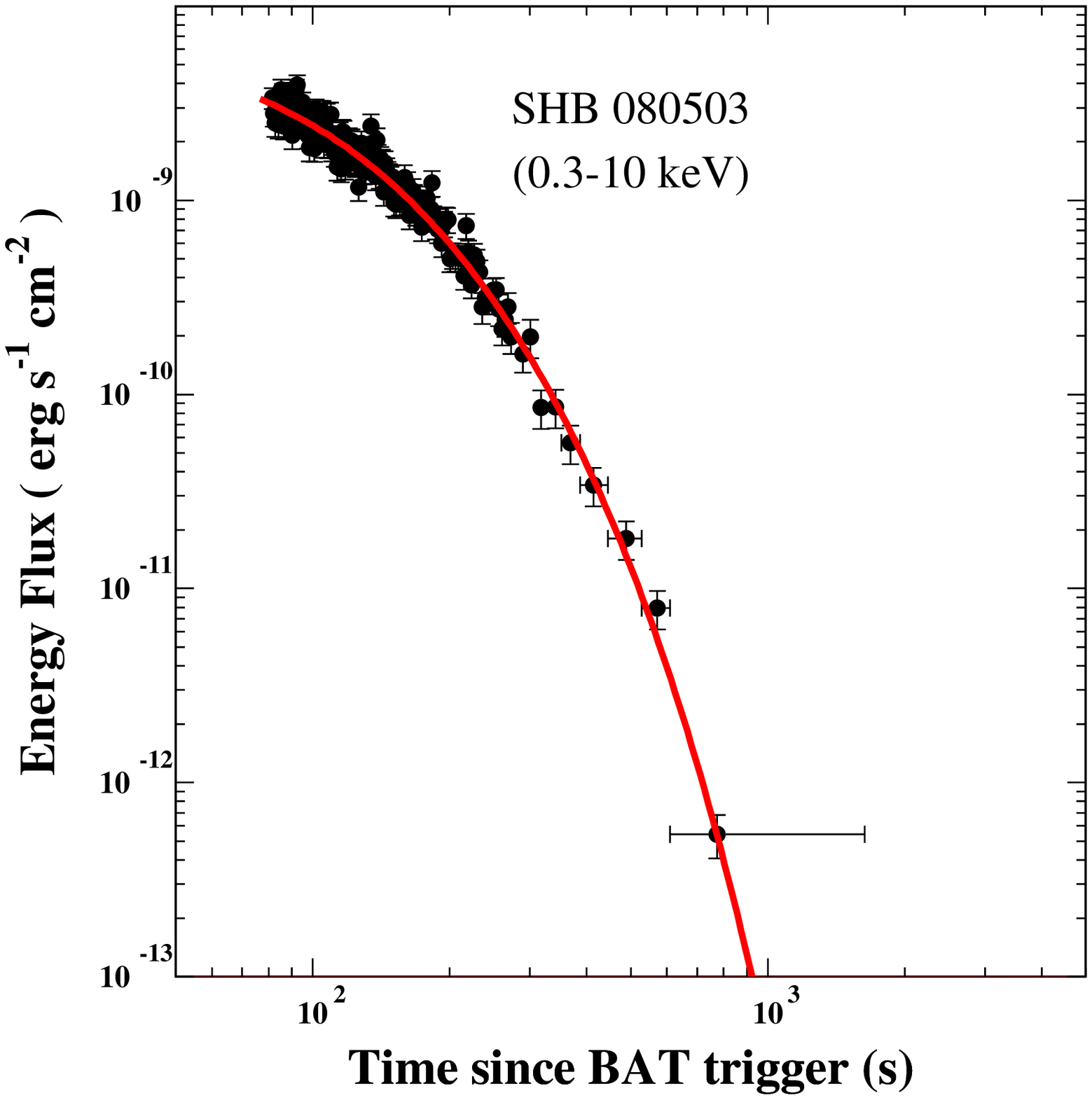,width=8.6cm,height=8.0cm}
\epsfig{file=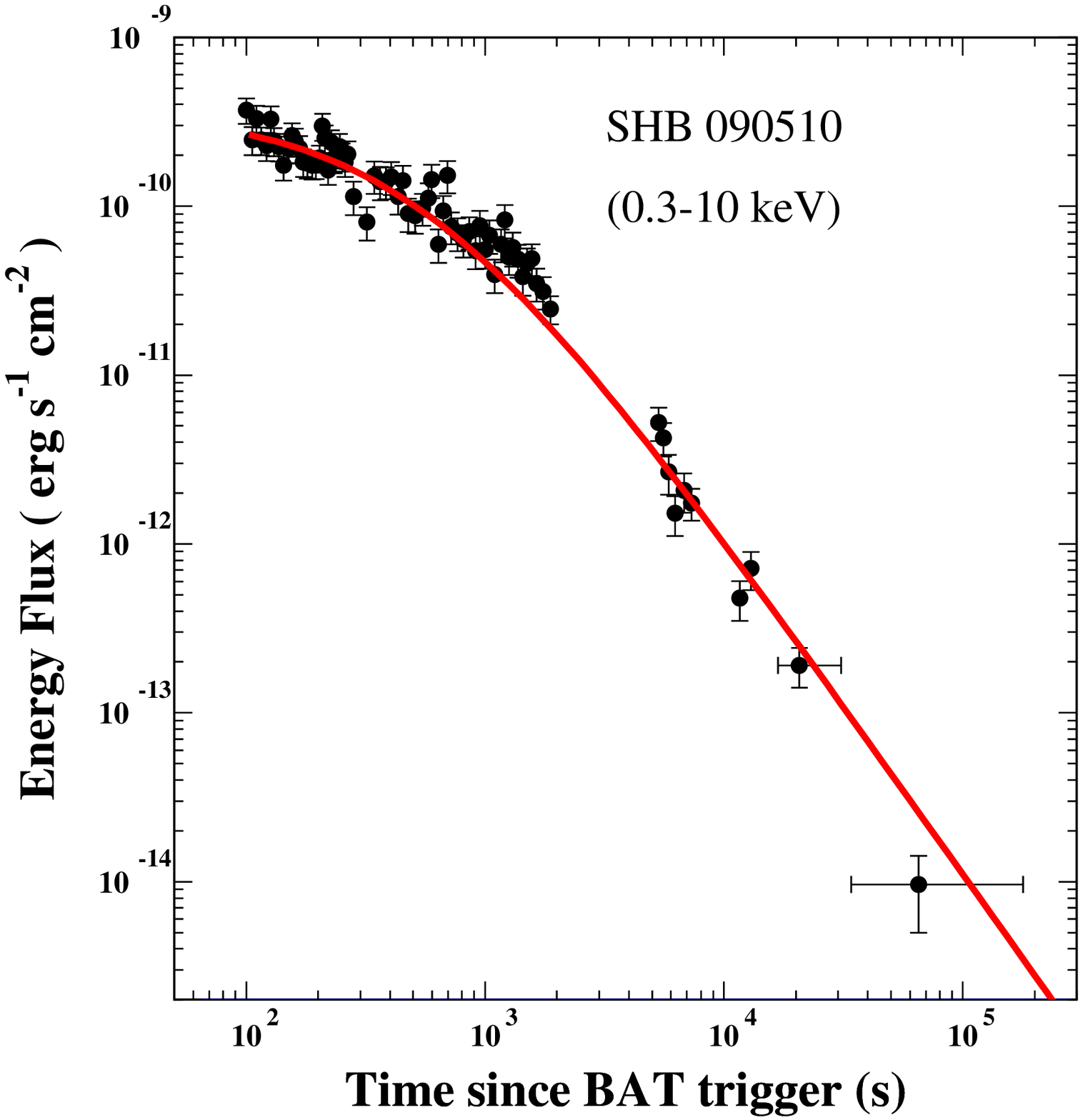,width=8.6cm,height=8.0cm}
}}
\vbox{
\hbox{
\epsfig{file=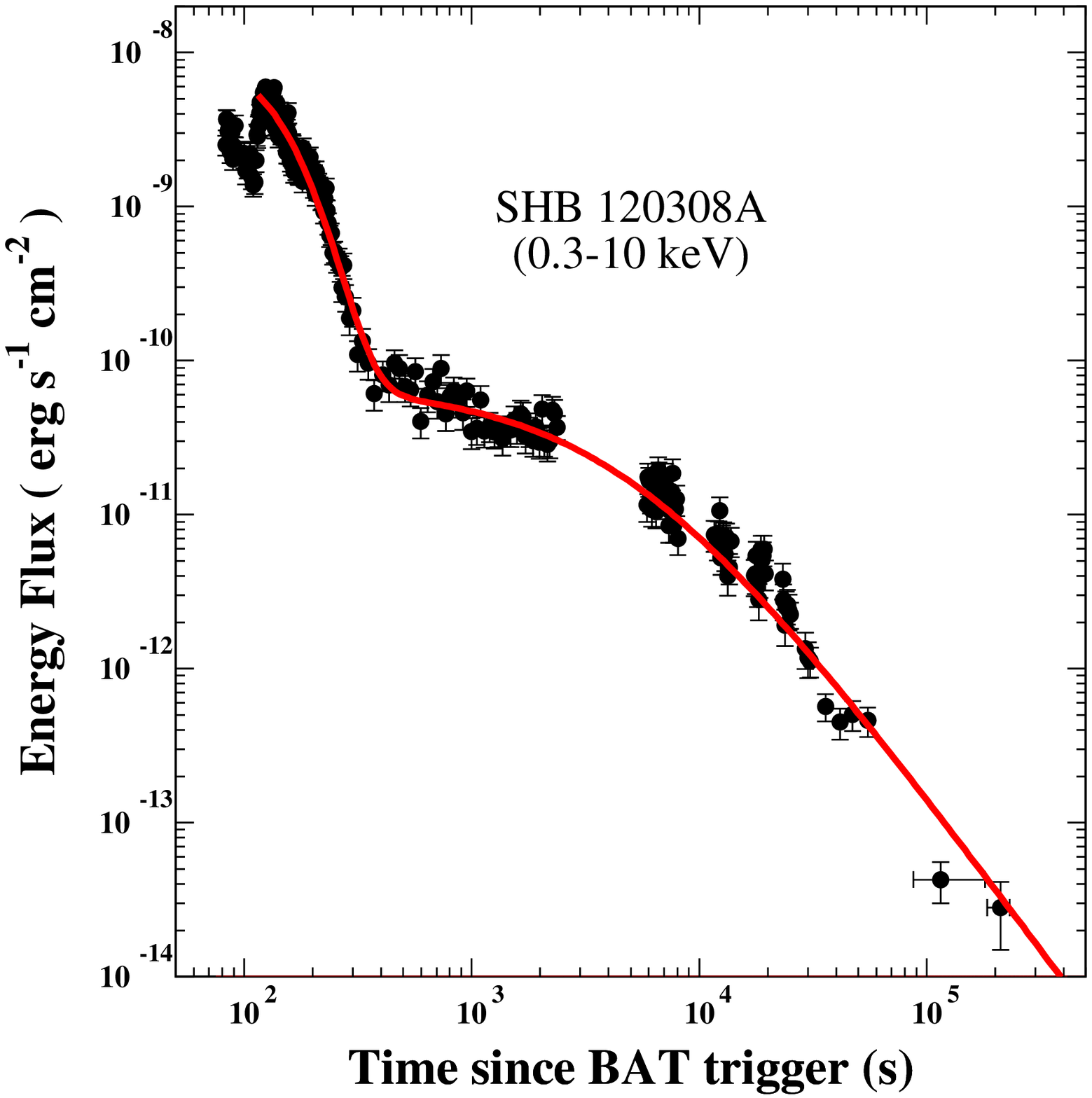,width=8.6cm,height=8.0cm}
\epsfig{file=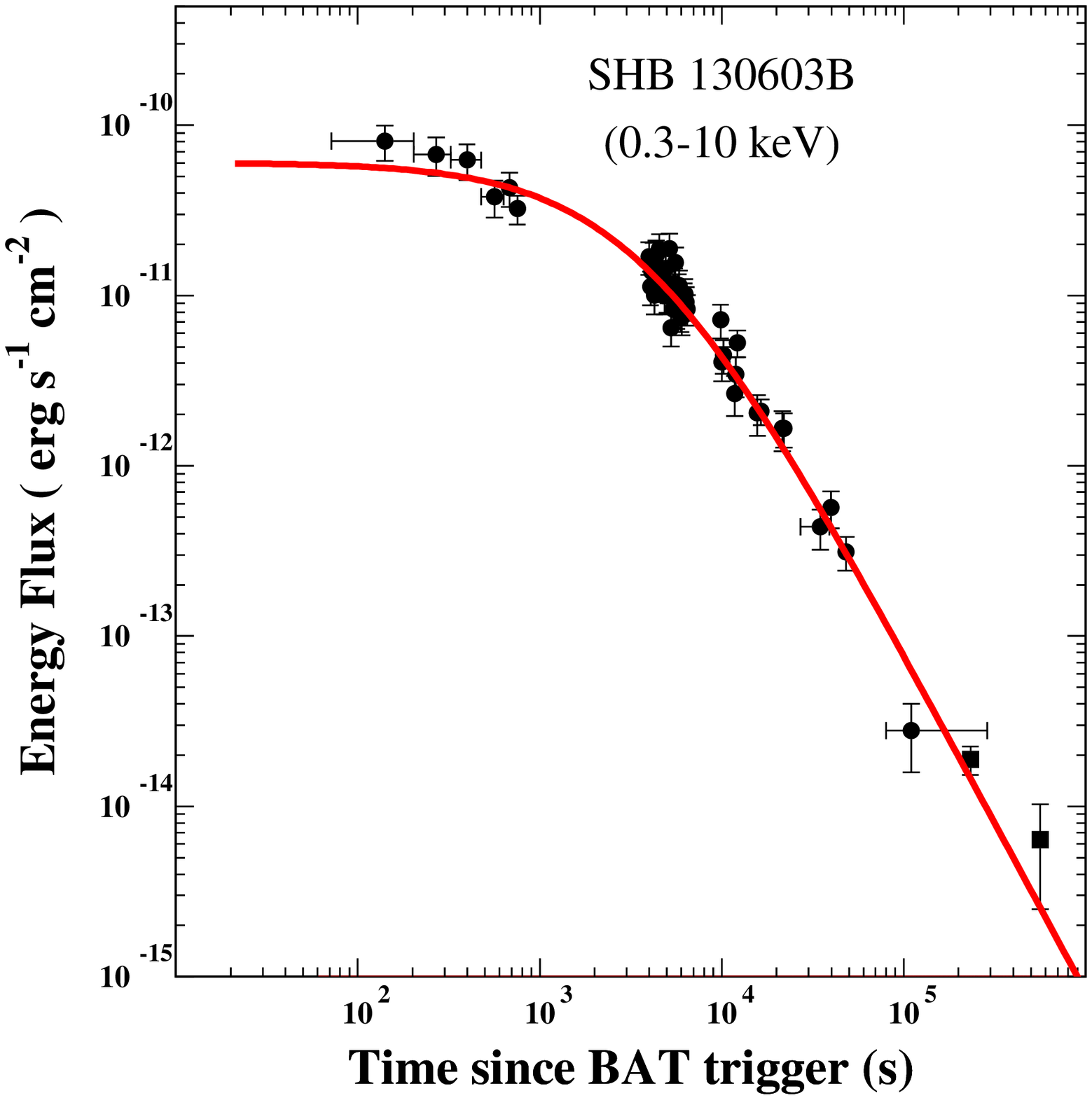,width=8.6cm,height=8.0cm}
}}
\vbox{
\hbox{
\epsfig{file=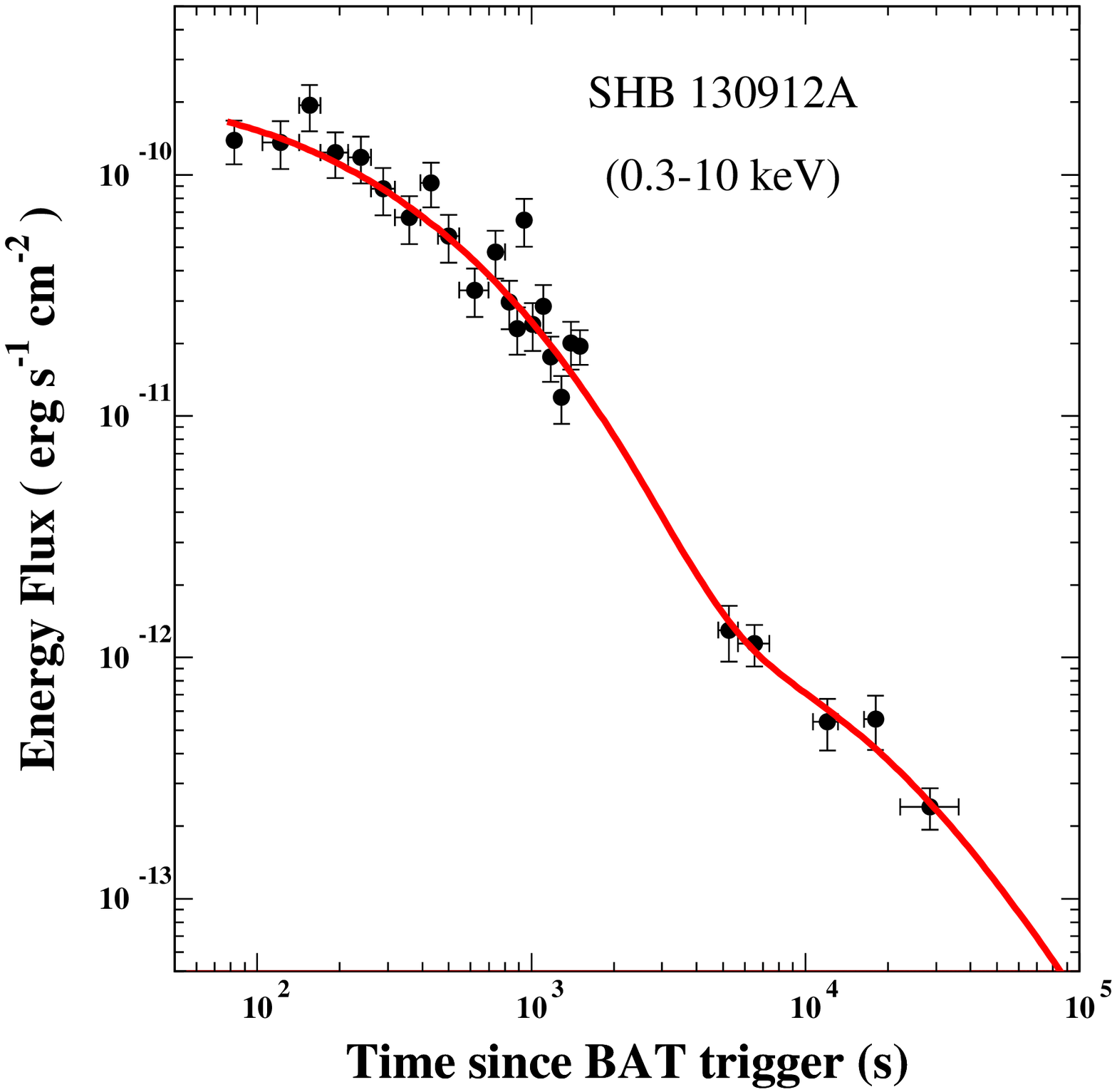,width=8.6cm,height=8.0cm}
\epsfig{file=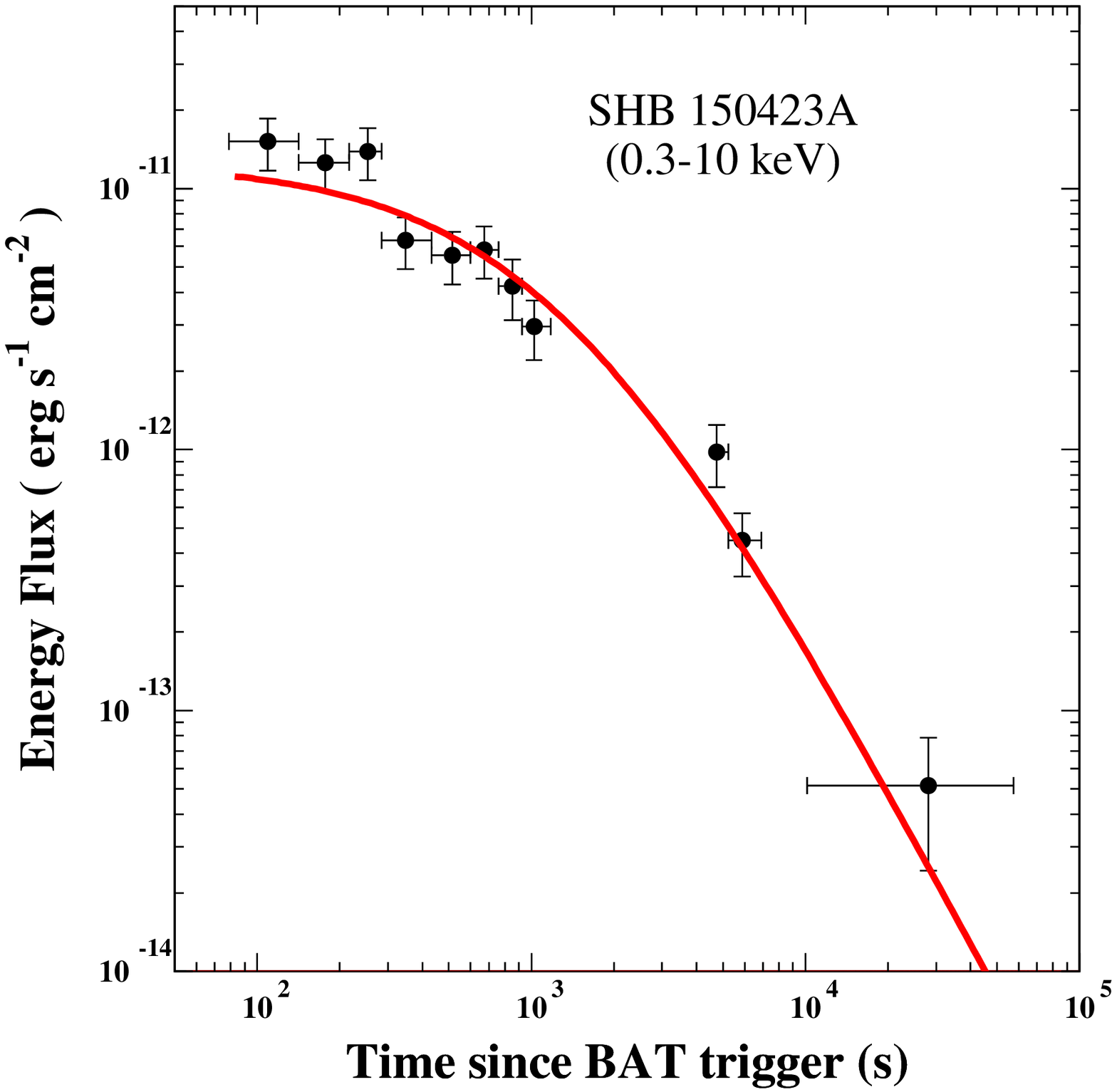,width=8.6cm,height=8.0cm}
}}
\caption{The X-ray light curve of SHBs
080503, 090510, 120308A, 130603, 130912A, 150423A
reported in the Swift-XRT GRB light curve repository [16]
and their best fit light curves assuming  ICS of GC light by a
highly relativistic jet  taken  over
by MSP afterglow, as given by by Eq.(4) or Eq.(10)
with the 2 or 4 parameters, respectively, listed in Table I.}
\label{F8}
\end{figure}

\begin{figure}[]
\centering
\vspace{-1cm}
\vbox{
\hbox{
\epsfig{file=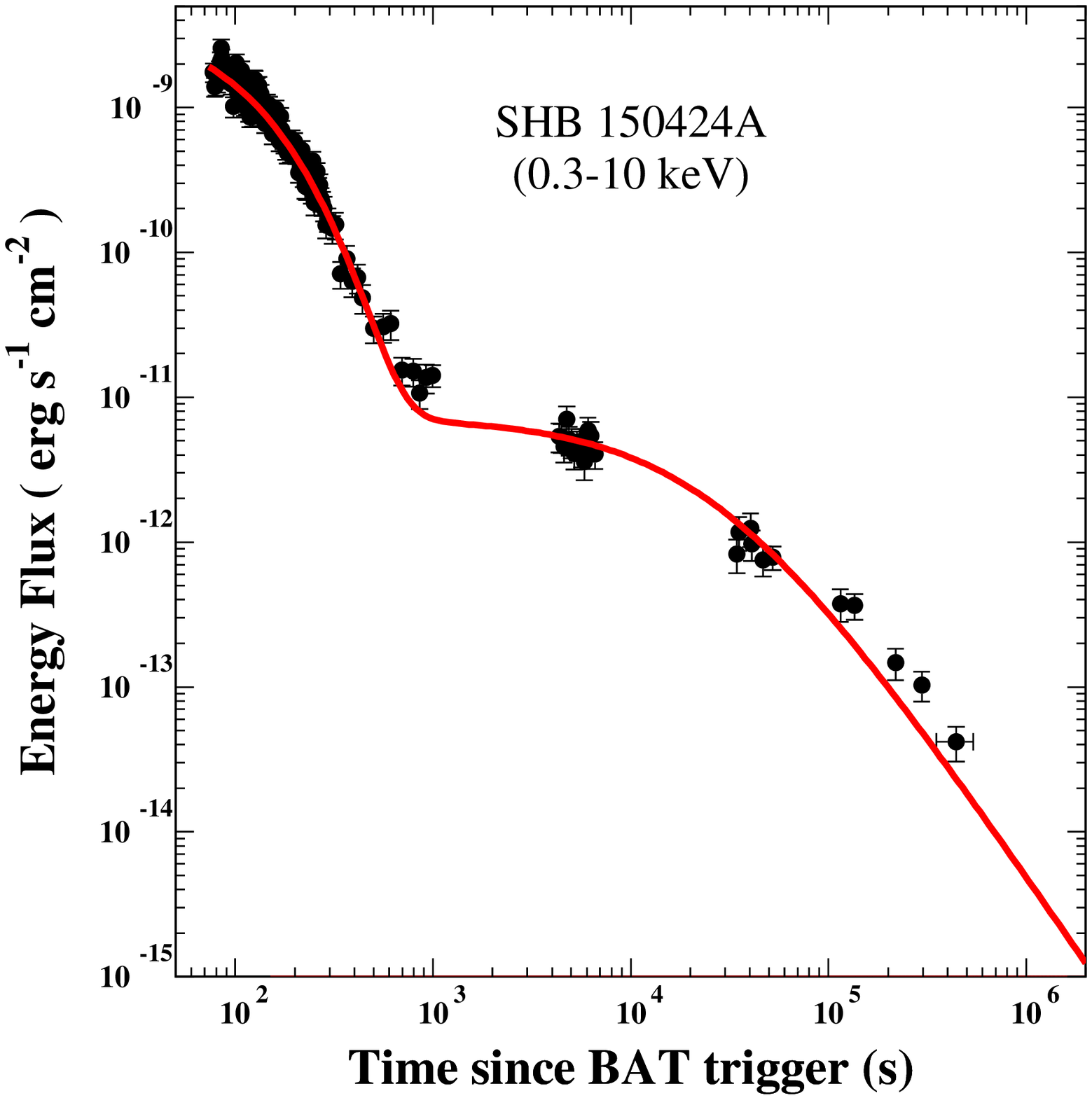,width=8.6cm,height=8.0cm}
\epsfig{file=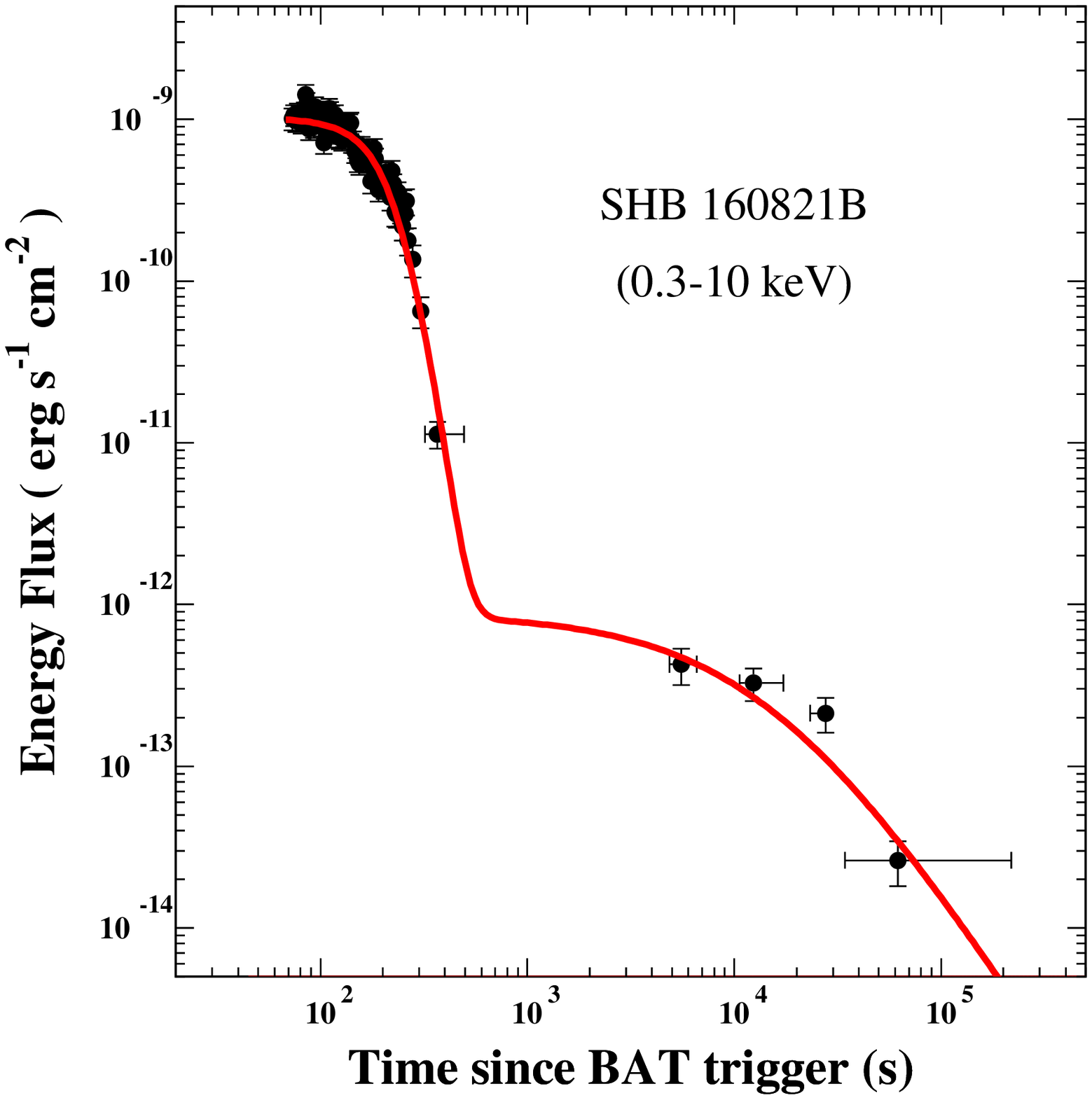,width=8.6cm,height=8.0cm}
}}
\caption{The X-ray light curve of  SHBs 150424A and  160821B
reported in the Swift-XRT GRB light curve repository [16]
and their best fit light curves assuming  ICS of GC light by a
highly relativistic jet  taken  over
by MSP afterglow, as given by by Eq.(4) or Eq.(10)
with the 2 or 4 parameters, respectively, listed in Table I.}
\label{F9}
\end{figure}

\begin{table*}
\caption{The best fit values of the parameters of Eq.(10) obtained  
from  the 20 well-sampled X-ray afterglows of SHBs 
measured with the Swift-XRT [14] and shown in Figures. 6-9 [16]. 
Also 
listed are the values of the period and polar magnetic field  of 
the MSPs at birth, as estimated from $F_p$ and $t_b$ for those SHBs 
with secured identity and redshift.}
\label{table1}
\centering    

\begin{tabular}{l l l l l l l l l l l}
\hline
\hline
~~SHB~~~&~~z~~~&~~~$F_{ee}$~~&~a~&~$t_{ee}$~&~~~$F_p$~&~$t_b$&$\chi^2/dof$&~~P~~\\ 
        &     &$[erg/s\,cm^2\,]$&        &   ~[s]~  &$[erg/s\,cm^2]$~& ~[s]~&   & [ms]\\   
\hline
051210  &      &  1.04E-9 & 1.55 & 113  &          &      &~~1.62~&      \\
051221A &0.5465&  3.70E-8 & [0]  & 570  & 2.59E-12 &42522 &~~1.44 & 3.53 \\
051227  &0.8   &  1.24E-9 & 1.33 & 103  & 6.88E-12 &~3829 &~~0.95 & 1.39 \\               
060313  &      & 2.66E-10 & 1.4  & 199  & 2.54E-11 &~4932 &~~0.89 &      \\
? 060614&0.125 & 1.64E-7  & [0]  & 111  & 1.08E-11 &49830 &~~1.39 & 48.0 \\
060801  &      & 1.32E-10 & 23.7  & 104 &         &       &~~1.23 &      \\
061006  &      & 4.14E-9  & [0]  &1222  & 8.46E-13 &33318 &~~1.64 &      \\
? 070714B &0.92~~& 1.8E-9   & 3.16 & ~76& 1.03E-10 &~~736 &~~1.41 & 6.50 \\        
070724A &0.457 & 2.11E-9  & 3.16 & ~67  & 1.06E-12 &19381 &~~4.68 & 6.31 \\
070809  &0.2187& 2.26E-11 & 1.18 & 183  & 3.14E-12 &17512 &~~1.22 & 7.82 \\
071227  &0.383~& 7.02E-10 &  80 & ~51   & 3.86E-12 &~3079 &~~1.97 & 3.98 \\
080123  &0.495~(?)&1.38E-9 & 199 & ~43  & 8.20E-13 &~5536 &~~1.93&       \\
080503  &      &  7.74E-9 & 1.25 & ~99  &          &      &~~1.28 &      \\
090510  &0.903~&          &      &      & 3.74E-10 &~~546 &~~1.83 & 7.04 \\ 
120308A &      &~9.94E-9  & 16   & ~56  & 6.87E-11 &4730  &~~1.51 &      \\
130603B &0.3564&          &      &      & 6.06E-11 &~3668 &~~1.19 & 1.05 \\            
130912A &      & 2.31E-10 & [0]  & 1691 & 1.72E-12 &17489 &~~1.47 &      \\
150423A &1.394~&          &      &      & 1.26E-11 &~1312 &~~1.35 & 0.60 \\
150424A&$1\pm.2$&4.21E-9  & [0]  & ~140  & 7.25E-12 &26524&~~1.53 & 1.08 \\
160821B &0.16 (?)& 1.03E-9& 118   & ~53  & 8.75E-13 &15295 &~~1.38&      \\
\hline  

\end{tabular}
\end{table*}

\end{document}